# Incorporation of prior knowledge of the signal behavior into the reconstruction to accelerate the acquisition of MR diffusion data

*Juan F. P. J. Abascal[1,2,3], Manuel Desco[1,2,4] and Juan Parra-Robles[1,2]*

[1] Departamento de Bioingeniería e Ingeniería Aeroespacial, Universidad Carlos III de Madrid, Madrid, Spain

[2] Instituto de Investigación Sanitaria Gregorio Marañón (IiSGM), Madrid, Spain

[3] Univ.Lyon, INSA-Lyon, Université Claude Bernard Lyon 1, UJM-Saint Etienne, CNRS, Inserm, CREATIS UMR 5220, U1206, Lyon, France

[4] Centro de Investigación en Red de Salud Mental (CIBERSAM), Madrid, Spain

[*] Corresponding author: E-mail: jmparra@hggm.es

## Abstract

Diffusion MRI measurements using hyperpolarized gases are generally acquired during patient breath hold, which yields a compromise between achievable image resolution, lung coverage and number of b-values. In this work, we propose a novel method that accelerates the acquisition of MR diffusion data by undersampling in both spatial and b-value dimensions, thanks to incorporating knowledge about the signal decay into the reconstruction (SIDER). SIDER is compared to total variation (TV) reconstruction by assessing their effect on both the recovery of ventilation images and estimated mean alveolar dimensions (MAD). Both methods are assessed by retrospectively undersampling diffusion datasets of normal volunteers and COPD patients (n=8) for acceleration factors between x2 and x10. TV led to large errors and artefacts for acceleration factors equal or larger than x5. SIDER improved TV, presenting lower errors and histograms of MAD closer to those obtained from fully sampled data for accelerations factors up to x10. SIDER preserved image quality at all acceleration factors but images were slightly smoothed and some details were lost at x10. In conclusion, we have developed and validated a novel compressed sensing method for lung MRI imaging and achieved high acceleration factors, which can be used to increase the amount of data acquired during a breath-hold. This methodology is expected to improve the accuracy of estimated lung microstructure dimensions and widen the possibilities of studying lung diseases with MRI.

# Introduction

The increase of MRI sensitivity through the use of hyperpolarized contrast media has enabled the development of imaging techniques to assess anatomical features and functional processes beyond the limits of conventional MRI [Salerno 2001, Ross 2015]. In particular, hyperpolarized noble gas MRI can provide quantitative maps of clinically relevant anatomical and physiological parameters (e.g. ventilation distribution [Deninger 2002], [Horn 2014], acinar airway dimensions [Yablonskiy 2009], alveolar oxygen partial pressure (pO2) [Deninger 1999, Marshall 2014], gas washout [Deppe 2011]). An increase in MRI signal is achieved through the use of laser polarization techniques (e.g. optical pumping-spin exchange [Walker 1997], [Parnell 2010] that result in a non-equilibrium net magnetization that is up to 5 order of magnitude higher than in conventional (thermally polarized) MRI.

However, the non-renewable nature of the magnetization in hyperpolarized gases imposes limits on the duration of MR image acquisition. In presence of oxygen, the T1 of hyperpolarized gases (15-20s) is of the order of duration of the breathhold that can be achieved by patients within the scanner. As a consequence of this limitation, together with high cost of the gas (e.g., $^4$He and isotopically enriched xenon), most hyperpolarized gas methods aim to perform a complete acquisition during a single breath-hold using a single hyperpolarized gas dose. Furthermore, there are advantages in the acquisition with more than one sequence or even different nuclei during a single breath-hold [Wild 2013]. This need for rapid acquisition has been addressed using different accelerated acquisition approaches, including parallel imaging [Lee 2006, [Chang 2015], and compressed sensing. Compressed sensing (CS) has been suggested for accelerating acquisition for hyperpolarized gas MRI [Ajraoui 2010], [Ajraoui 2013]. In these approaches the acquisition was accelerated in the spatial encoding direction and the image reconstructed using spatial total variation (TV). More recently, [Chan 2016] have used compressed sensing to acquire diffusion images of hyperpolarized gases in the lungs. Diffusion images are sensitive to changes in lung microstructure [Swift 2005] due to disease, and are used to estimate the dimensions of acinar airways using theoretical models obtained from numerical simulations [Yablonskiy 2009]. These theoretical models require the acquisition of images for several diffusion sensitization values (b-values), which together with the longer duration of the diffusion scan (due to the presence of diffusion gradients) result on long acquisition times. Due to the limitation by the breath hold duration, there is a compromise between achievable image resolution, number of slices and number of b-values, thus limiting the accuracy and number of parameters of the theoretical models [Parra-Robles 2012a]. Typically, most implemented protocols acquire 5 slices (10 mm thick, spacing 10 mm) with 64x64 pixels resolution and 4-6 b-values, hence sacrificing lung coverage [Parra-Robles 2012b]. Chan et al. achieved full lung coverage by using compressed sensing with a 3D diffusion acquisition and undersampling along both spatial directions [Chan 2015].

Although diffusion images are more sparse in the b-direction than in the spatial domain, the feasibility of exploiting sparsity along both spatial encoding and b-value directions has not been studied in hyperpolarized gas MRI. In other MR applications, such as in cardiac cine MRI [Lingala 2011], [Montesinos 2013], [Abascal 2014], fMRI [Chiew 2015], [Chavarrias 2015], and diffusion tensor MRI [Landman 2012], [Ning 2016], among others, CS has led to large acceleration factors by exploiting high data dimensionality.

In this work, we propose a novel compressed sensing method that incorporates a model of the signal decay as prior information into the reconstruction (SIDER), to accelerate the acquisition of MR diffusion data by undersampling in both spatial and b-value dimensions. We incorporate the knowledge of the diffusion signal behavior into the reconstruction to accelerate the acquisition of MR diffusion data. The proposed method is compared to TV and zero filling reconstructions by assessing its effect on the estimated parameters of a stretched exponential model, which has been used to estimate mean alveolar dimensions [Parra-Robles 2014]. Methods were assessed on control and COPD patient data sets (n=8 in total) using retrospective undersampling simulations, adopting as gold standard the fully sampled data.

## Methods

### Image reconstruction methods

#### Total variation

Previous compressed sensing studies for MRI using hyperpolarized gases assumed that each ventilation image $u_i$ is sparse under a transformation $\Psi$, which accounts for spatial sparsity [Ajraoui 2010], [Ajraoui 2013], [Chan 2015]. The most common choice for $\Psi$ is the gradient that leads to the total variation functional. If $F$ represents the undersampled Fourier transform and $f_i$ represents the undersampled k-space corresponding to the $i$-th b-value, then the total variation problem is given by

$$\min_{u_i} \|\nabla u_i\|_1 \text{ such that } \|Fu_i - f_i\|^2 \leq \sigma^2 \quad , \tag{1}$$

where $\nabla=(\nabla_x,\nabla_y)$, $i=1,\ldots,B$, and $B$ is the total number of b-values.

### Signal decay based reconstruction method

We propose a novel compressed sensing method that incorporates a model of the signal decay into the reconstruction (SIDER). It combines TV with a penalty function that promotes sparsity across the b-direction as follows:

$$\min_{u} \alpha \|\nabla u\|_1 + \beta \|Mu\|_1 \text{ such that } \|Fu - f\|^2 \leq \sigma^2, \quad (2)$$

where $u$ and $f$ are the ventilation images and undersampled data corresponding to all values of $b$, $u=[u_1,\ldots,u_B]$, $f=[f_1,\ldots,f_B]$, $F$ is the undersampled multislice Fourier transform, and M is an operator that encodes the relationship between ventilation images for consecutives values of $b$. This relationship can be approximated using a stretched exponential model [Parra-Robles 2014], [Chan 2015] as

$$Mu_j = u_j(b_j) - u_{j-1}(b_{j-1}) \exp\left[-\left(\left(\bar{D}b_j\right)^{\bar{\alpha}} - \left(\bar{D}b_{j-1}\right)^{\bar{\alpha}}\right)\right], \quad (3)$$

where $\bar{D}$ and $\bar{\alpha}$ are estimated average value of diffusivity and heterogeneity index, respectively, which can be obtained from a previous reconstructed image (here we used the image provided by TV method).

### Split Bregman formulation

Problems in Eq. (1) and Eq.(2) were solved using the Split Bregman formulation, which efficiently handles L1-based constrained problems [Osher 2005], [Goldstein 2009], [Montesinos 2013], [Abascal 2014]. Using this formulation, constrained problems are converted to equivalent unconstrained problems, where constraints are imposed iteratively using the Bregman iteration. L2- and L1-norm functionals are separated into several subproblems, which are solved analytically in alternating steps. The subproblem including the L2-norm functionals results in a linear system that can be efficiently solved using iterative Krylov solvers and subproblems including L1-norm functionals are solved using shrinkage formulas. As TV can be obtained from SIDER by making β=0, we develop the formulation for the general case of SIDER.

To perform the split, we include the new variables $d_x$, $d_y$, and $w$ and formulate a new problem that is equivalent to Eq. (2)

$$\min_{u, d_x, d_y, w} \alpha \|(d_x, d_y)\|_1 + \beta \|w\|_1 \text{ such that } \|Fu - f\|^2 \leq \sigma^2,$$
$$d_x = \nabla_x u, \quad d_y = \nabla_y u, \quad w = Mu \quad (4)$$

Eq. (4) is now easily managed using an equivalent unconstrained optimization approach where constraints are imposed by adding Bregman iterations $b_i$,

$$\min_{u, d_x, d_y, w} \alpha \left\|(d_x, d_y)\right\|_1 + \beta \|w\|_1 + \frac{\mu}{2}\|Fu - f^k\|_2^2 + \frac{\lambda}{2}\|d_x - \nabla_x u - b_x^k\|_2^2 +$$
$$\frac{\lambda}{2}\|d_y - \nabla_y u - b_y^k\|_2^2 + \frac{\lambda}{2}\|w - Mu - b_w^k\|_2^2 \tag{5}$$

where μ is a regularization parameter that weights the data fidelity term, λ is another regularization parameter that weights the terms imposing the constraints for the dummy variables, $k$ is the iteration number and the Bregman iterations are updated as

$$\begin{aligned}
b_x^{k+1} &= b_x^k + \nabla_x u^{k+1} - d_x^{k+1} \\
b_y^{k+1} &= b_y^k + \nabla_y u^{k+1} - d_y^{k+1} \\
b_w^{k+1} &= b_w^k + Mu^{k+1} - w^{k+1} \\
f^{k+1} &= f^k + f - Fu^{k+1}
\end{aligned} \tag{6}$$

Since $u$ and auxiliary variables $w$, $d_x$, and $d_y$ are independent of each other, Eq. (5) can now be split into several equations (one for each variable) that are solved sequentially, as follows:

$$\begin{aligned}
u^{k+1} &= \min_u \frac{\mu}{2}\|Fu - f^k\|_2^2 + \frac{\lambda}{2}\|d_x^k - D_x u - b_x^k\|_2^2 + \frac{\lambda}{2}\|d_y^k - D_y u - b_y^k\|_2^2 \\
&\quad + \frac{\lambda}{2}\|w^k - Mu - b_w^k\|_2^2 \\
(d_x^{k+1}, d_y^{k+1}) &= \min_{d_x, d_y} \alpha \|(d_x, d_y)\|_1 + \frac{\lambda}{2}\|d_x - D_x u^{k+1} - b_x^k\|_2^2 + \frac{\lambda}{2}\|d_y - D_y u^{k+1} - b_y^k\|_2^2 \\
w^{k+1} &= \min_w \beta \|w\|_1 + \frac{\lambda}{2}\|w - Mu^{k+1} - b_w^k\|_2^2
\end{aligned} \tag{7}$$

Since the solution of $u$ only involves L2-norm functionals, it can be obtained exactly as the solution of the linear system

$$\begin{aligned}
Ku^{k+1} &= r^k \\
K &= \mu F^T F + \lambda D_x^T D_x + \lambda D_y^T D_y + \lambda M^T M \\
r^k &= \lambda D_x^T (d_x^k - b_x^k) + \lambda D_y^T (d_y^k - b_y^k) + \lambda M^T (w^k - b_w^k)
\end{aligned} \tag{8}$$

Note that Eq. (8) constitutes a very large-scale problem, where K=NxN and N is the number of pixels, yet it can be solved efficiently using a Krylov solver, such as the biconjugate gradient stabilized method, which involves only matrix-vector multiplications:

$$\mu F^T (Fu) + \lambda D_x^T (D_x u) + \lambda D_y^T (D_y u) + \lambda M^T (Mu) = r^k \tag{9}$$

The auxiliary variables $d_x$, $d_y$, and $w$ are solved analytically using shrinkage formulas, which are thresholding operations (1,2).

$$d_j^{k+1} = \max\left(s^k - \alpha/\lambda, 0\right)\frac{\left|D_j u^{k+1} + b_j^k\right|}{s^k}, j = x, y$$

$$s^k = \sqrt{\left|D_x u^{k+1} + b_x^k\right|^2 + \left|D_y u^{k+1} + b_y^k\right|^2},$$

$$w^{k+1} = \text{shrink}\left(Mu^{k+1} + b_w^k, \beta/\lambda\right) =$$

$$= \max\left(\left|Mu^{k+1} + b_w^k\right| - \beta/\lambda, 0\right) \text{sign}\left(Mu^{k+1} + b_w^k\right)$$

(10)

**Selection of the reconstruction algorithm parameters**

Regularization parameters related to the Bregman iterations in TV and SIDER methods (µ and λ in [Eq. (7)]) were selected following suggestions from previous studies [Goldstein 2009], [Abascal 2011]. The number of iterations was chosen to minimize the mean-square error considering the fully sampled image as the correct solution. For µ≤2 the method converged to the same solution at different iteration numbers. For µ>2 overfitting occurred at the first iterations and an optimal solution was not met. Increasing λ (higher weight to regularization penalty terms) lowers the threshold in the shrinkage formulas (Eq. (10)), resulting in slightly faster convergence, but for λ>20 the convergence became unstable. Values of λ<1 resulted in low smoothing at the first iterations and an excessively high threshold in the shrinkage formulas, which also led to unstable convergence and large errors. Values in the range 1≤λ≤20 led to similar results.

The weighting parameters that control the relative degree of sparsity between TV (α) and model decay sparsity (β) were heuristically determined as follows. Increasing α above one (higher sparsity to TV) led to images largely affected by cartoon-like artefacts and large solution errors. Selecting α<0.1 led to excessively fast convergence, compromising the robustness of the algorithm. Values of α in the range 0.1<α<1 suppressed most noise in the image and led to similar results. Increasing β imposed higher sparsity to the model of the signal decay than to TV, leading to lower solution errors, images with less cartoon-like artefacts and images less affected by noise (especially images corresponding to large values of b). However, the larger the value of β, the slower the convergence. We found that values in the range 0.2≤β≤1 were a good compromise. Table 1 shows a summary of the regularization parameter values used for both TV and SIDER methods.

Table 1. Regularization parameters selected for TV and SIDER methods.

|       | α   | β   | µ | λ | γ    |
|-------|-----|-----|---|---|------|
| **TV**    | 0   | 0   | 1 | 1 | 1    |
| **SIDER** | 0.2 | 0.2 | 1 | 1 | 0.01 |

## Datasets and retrospective undersampling

Eight fully sampled diffusion datasets were available from earlier work, three from normal volunteers and five from three patients with COPD (two patients had two acquisitions at different sessions) [Parra-Robles 2012b], [Parra-Robles 2014]. Data consisted of five slices (10 mm thick with 10 mm gap between slices), 64x64 resolution and 5 b-values (0, 1.6, 3.2, 4.8 and 6.4 s/cm$^2$), obtained with a diffusion time of 1.6 ms. These data were acquired in a GE HDx 1.5 T scanner (GE Healthcare, USA), using $^3$He gas polarized buy means of a SEOP commercial polarizer (Helispin, GE Healthcare, USA), that achieved polarizations of 30-40%.

These datasets were retrospectively undersampled to simulate CS acquisition and reconstruction. Quasi-random undersampling patterns were created [Lustig 2007], [Montesinos 2013], in which randomization was performed in the phase encoding direction and through the b-direction (Fig 1). This allowed us to exploit data redundancy in two dimensions. We analyzed the results for acceleration factors of x2, x4, x5, x7, and x10.

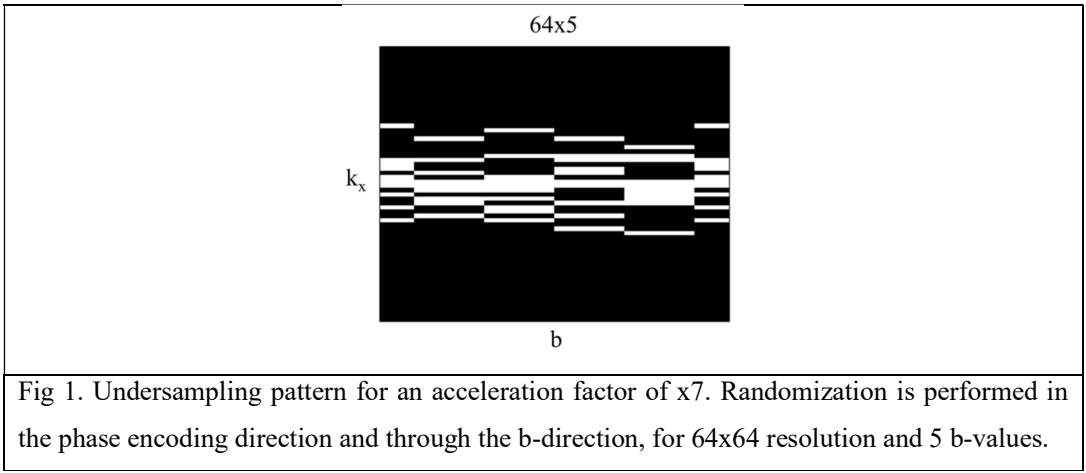

Fig 1. Undersampling pattern for an acceleration factor of x7. Randomization is performed in the phase encoding direction and through the b-direction, for 64x64 resolution and 5 b-values.

## Image analysis and evaluation

To evaluate the results, we first fitted the reconstructed signal $u(b)$, on a pixel-by-pixel basis, to the stretched exponential model, which estimates maps of the distributed diffusion coefficient $D$ and heterogeneity index α [Parra-Robles 2014]:

$$u(b) = u(0) \exp\left(-(bD)^\alpha\right). \tag{11}$$

As reconstructed images $u(b)$ are noisy, especially for patients and larger values of b, images $u(b)$ were smoothed using a Gaussian filter (window of three neighboring pixels and SD of one pixel) before fitting the model in Eq. (11). Estimation of $D$ and α was done only within a mask that had been created by segmenting ventilation images for fully sampled data. Then, we estimated the mean alveolar length, $L_m$, from $D$ and $\alpha$ as described in [Parra-Robles 2014], [Chan 2015]. $L_m$ was estimated only in the ranges $0<D<0.9$ and $0.3<\alpha<1.3$, which were assumed as physically reliable. The process for estimating $L_m$ from ventilation images, $u(b)$, as well as the differences between patient and control data sets are shown in Fig 2.

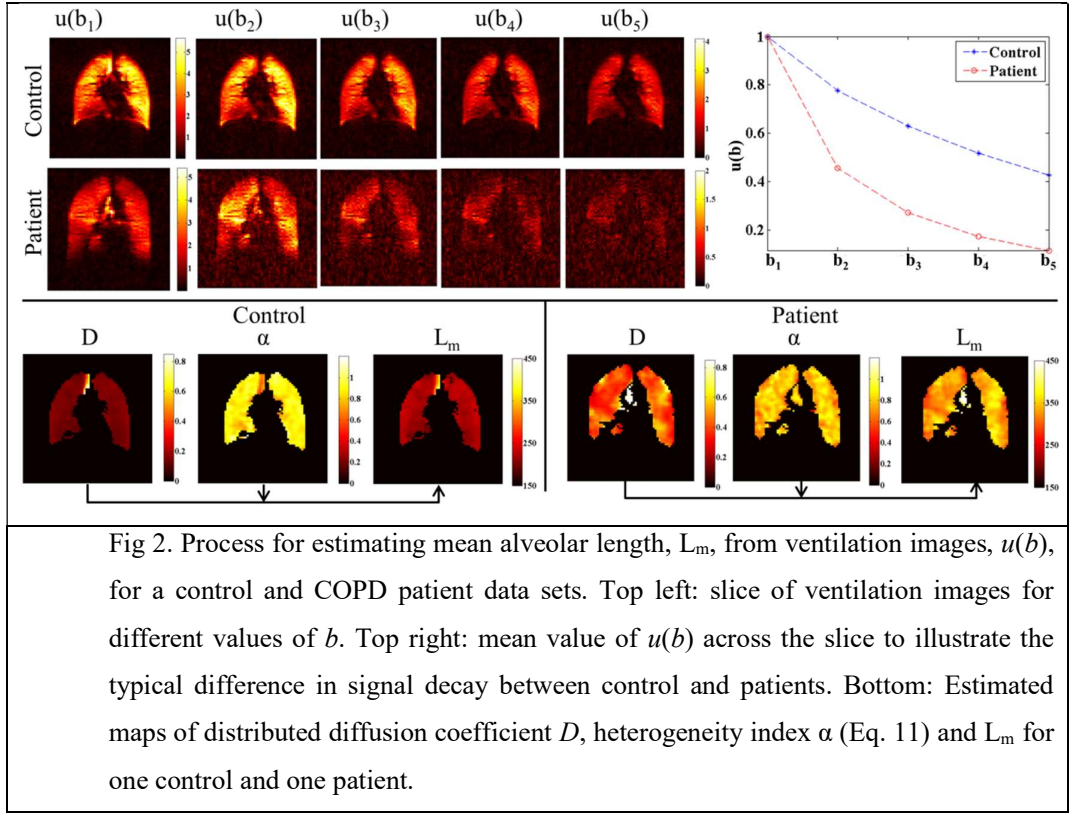

Fig 2. Process for estimating mean alveolar length, $L_m$, from ventilation images, $u(b)$, for a control and COPD patient data sets. Top left: slice of ventilation images for different values of $b$. Top right: mean value of $u(b)$ across the slice to illustrate the typical difference in signal decay between control and patients. Bottom: Estimated maps of distributed diffusion coefficient $D$, heterogeneity index $\alpha$ (Eq. 11) and $L_m$ for one control and one patient.

All retrospectively undersampled data sets (n=8 for acceleration factors x2, x4, x5, x7, and x10) were reconstructed with Zero filling (ZF), TV and SIDER methods. Methods were evaluated in terms of the following metrics: 1) relative MSE of the recovered ventilation images, 2) relative MSE, 3) histograms and 4) mean values of the estimated maps of mean alveolar length. MSE was computed by adopting as gold standard the images and maps obtained from the fully sampled data and results are given as mean and SD across all data sets. To assess the statistical significance of the difference between methods we used a Mann-Whitney test, as it is robust and avoids the assumption of normality in the data. Histograms and images are shown for one control and one patient data set for all methods. Mean and SE of $L_m$ in a region of interest (one slice) are shown for the three patients to verify that errors due to the undersampling were smaller than patient variability. Then, images and histograms for all data sets are shown for SIDER method and variations due to the undersampling were compared to intragroup differences in control and patient datasets.

# Results

## Comparison of methods

Fig 3 (left) shows the MSE of the reconstruction of ventilation images (for b=0) with the different methods for all the acceleration factors tested. SIDER method led to significantly lower MSE than ZF and TV in all cases (for b=0); SIDER also led to lower MSE for large values of b but differences were significant only at high accelerations (results not shown). Adopting an MSE of 10 % as a reference for the comparison, acceleration factors achieved by the different methods were x2 by ZF, x5 by TV and x10 by SIDER. SIDER also presented significantly lower MSE of mean alveolar length for all acceleration factors (Fig 3, right). Similarly to MSE of ventilation images, SIDER presented for x10 the same MSE than TV for x5.

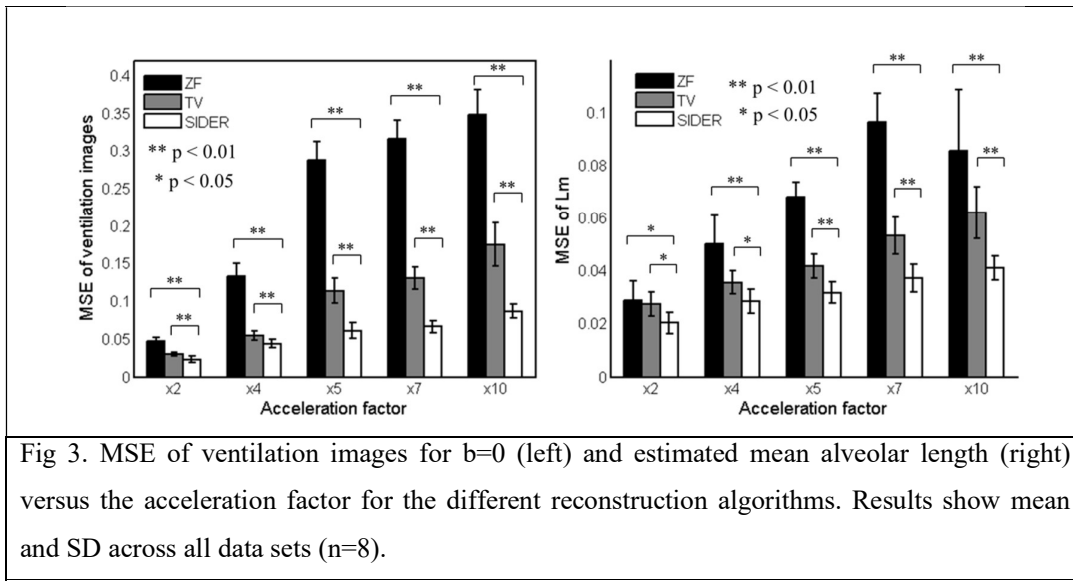

Fig 3. MSE of ventilation images for b=0 (left) and estimated mean alveolar length (right) versus the acceleration factor for the different reconstruction algorithms. Results show mean and SD across all data sets (n=8).

Maps of mean alveolar length for the different methods and for the highest acceleration factors (x5, x7, and x10) are shown in Figs 4 and 5 for one control and patient datasets, respectively. In the case of the control dataset, for acceleration factor x5, ZF and TV led to errors and artifacts while SIDER provided maps that were almost identical to the fully sampled data set. For acceleration factor x10, ZF and TV led to larger errors and artefacts and a shift in the mean value. On the contrary, SIDER still preserved image quality with small deviations in the estimated maps for x10.

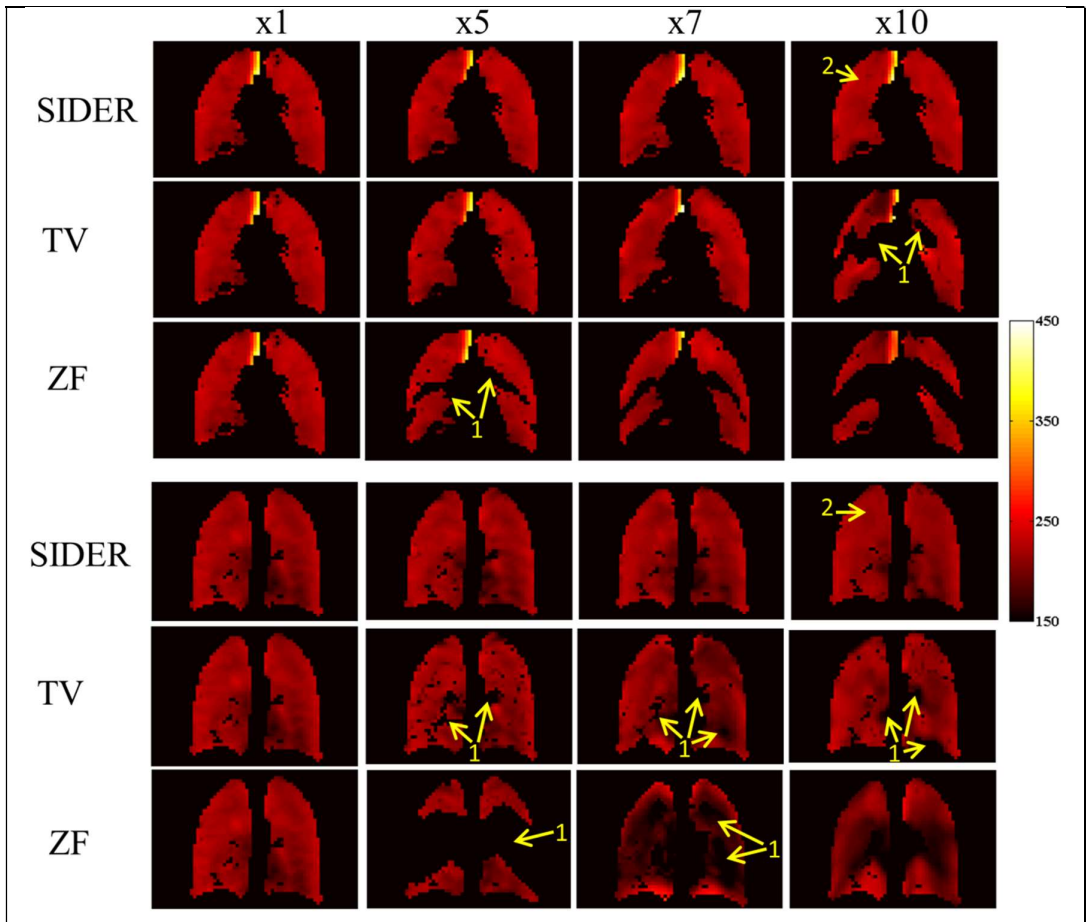

Fig 4. Two slices of estimated mean alveolar length maps for the different methods and acceleration factors, for a control data set. Arrows point to areas where errors are more visible: artefacts (1) and variations in the estimated maps (2). Videos of results for all data sets are available from https://github.com/HGGM-LIM/compressed-sensing-diffusion-lung-MRI.

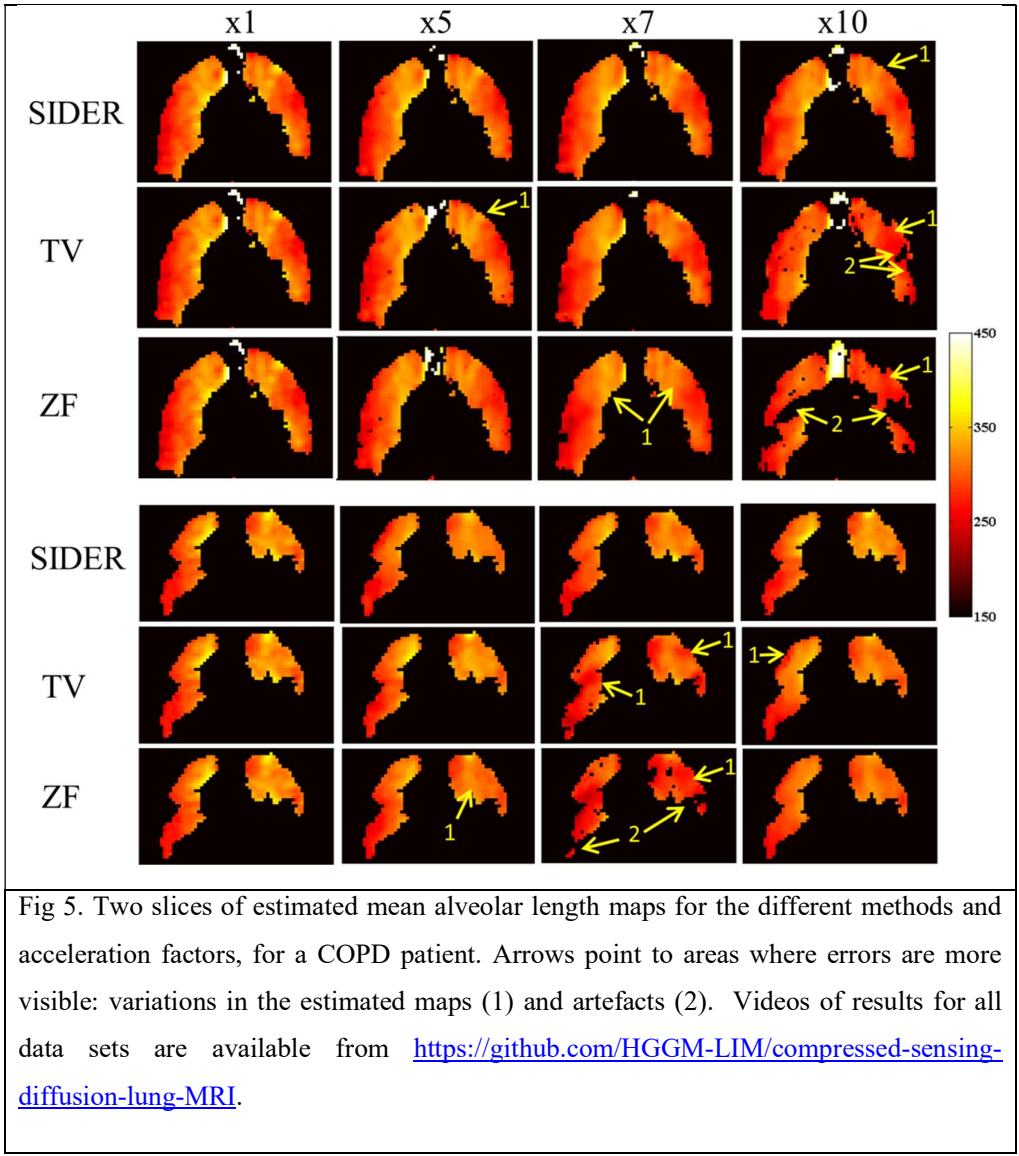

Fig 5. Two slices of estimated mean alveolar length maps for the different methods and acceleration factors, for a COPD patient. Arrows point to areas where errors are more visible: variations in the estimated maps (1) and artefacts (2). Videos of results for all data sets are available from https://github.com/HGGM-LIM/compressed-sensing-diffusion-lung-MRI.

Fig. 6 shows the histograms of estimated mean alveolar length from images reconstructed with the different algorithms for a control and patient data set. ZF and TV led to larger variations as the acceleration increased, becoming very noticeable above x7. On the contrary, SIDER presented histograms close to the target for accelerations up to x10.

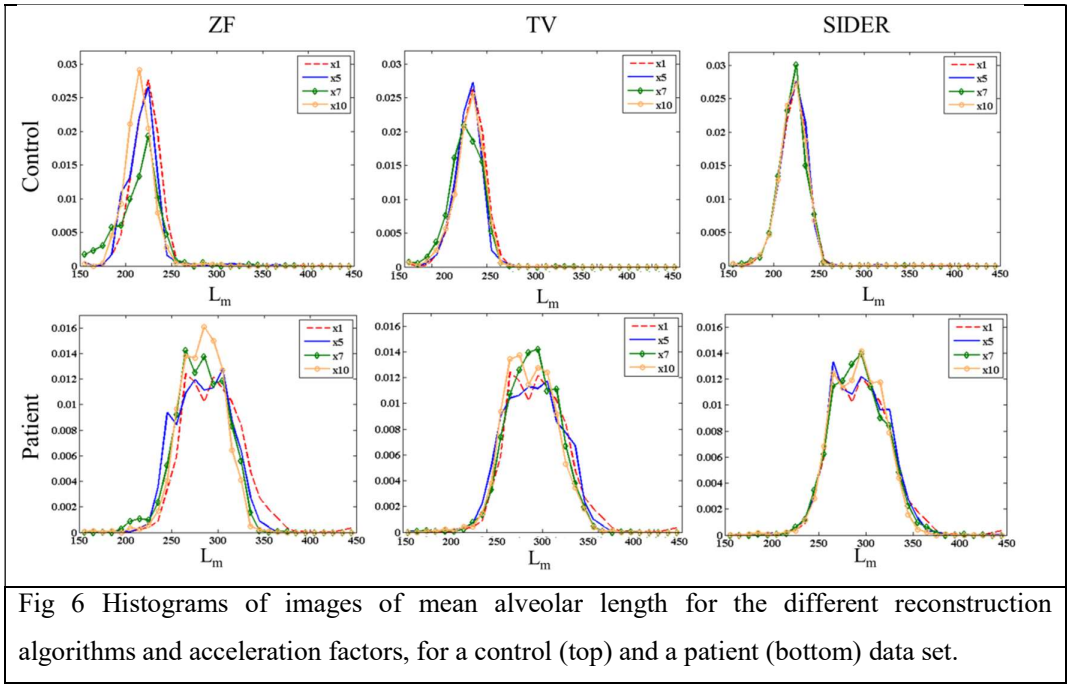

Fig 6 Histograms of images of mean alveolar length for the different reconstruction algorithms and acceleration factors, for a control (top) and a patient (bottom) data set.

Fig 7 shows mean and SE of mean alveolar length across one slice for three different patients. For accelerations larger than five errors by ZF and TV were larger than differences between data sets. On the contrary, SIDER led to small errors for all acceleration factors.

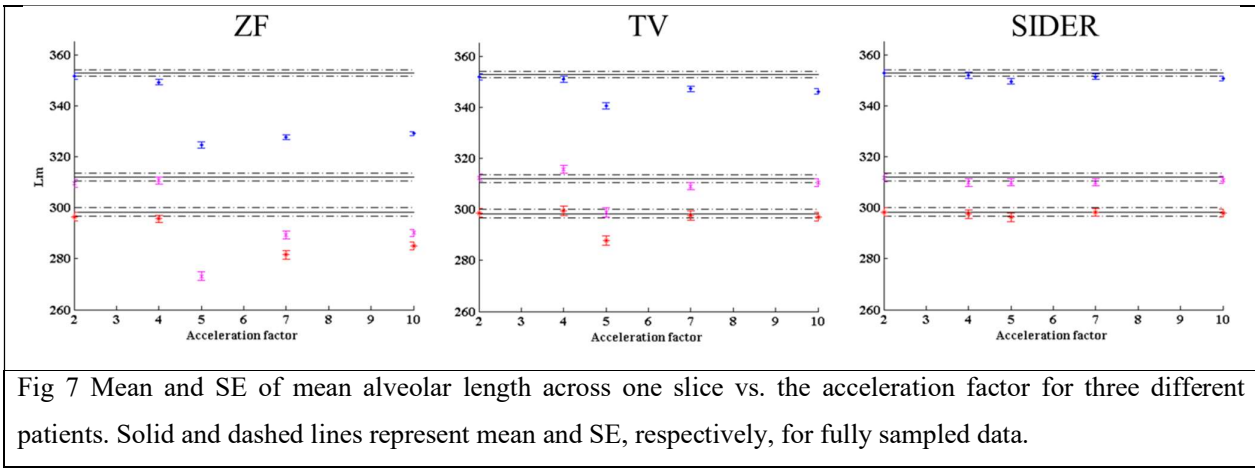

Fig 7 Mean and SE of mean alveolar length across one slice vs. the acceleration factor for three different patients. Solid and dashed lines represent mean and SE, respectively, for fully sampled data.

### Analysis of the acceleration factor in retrospective data

Fig 8 shows a slice of the map of estimated mean alveolar length obtained by the SIDER method for the rest of datasets (two controls and four patient data sets) and for the highest acceleration factors. Image quality was preserved for all accelerations, but there were small deviations from the fully sampled remained for an acceleration factor of x10.

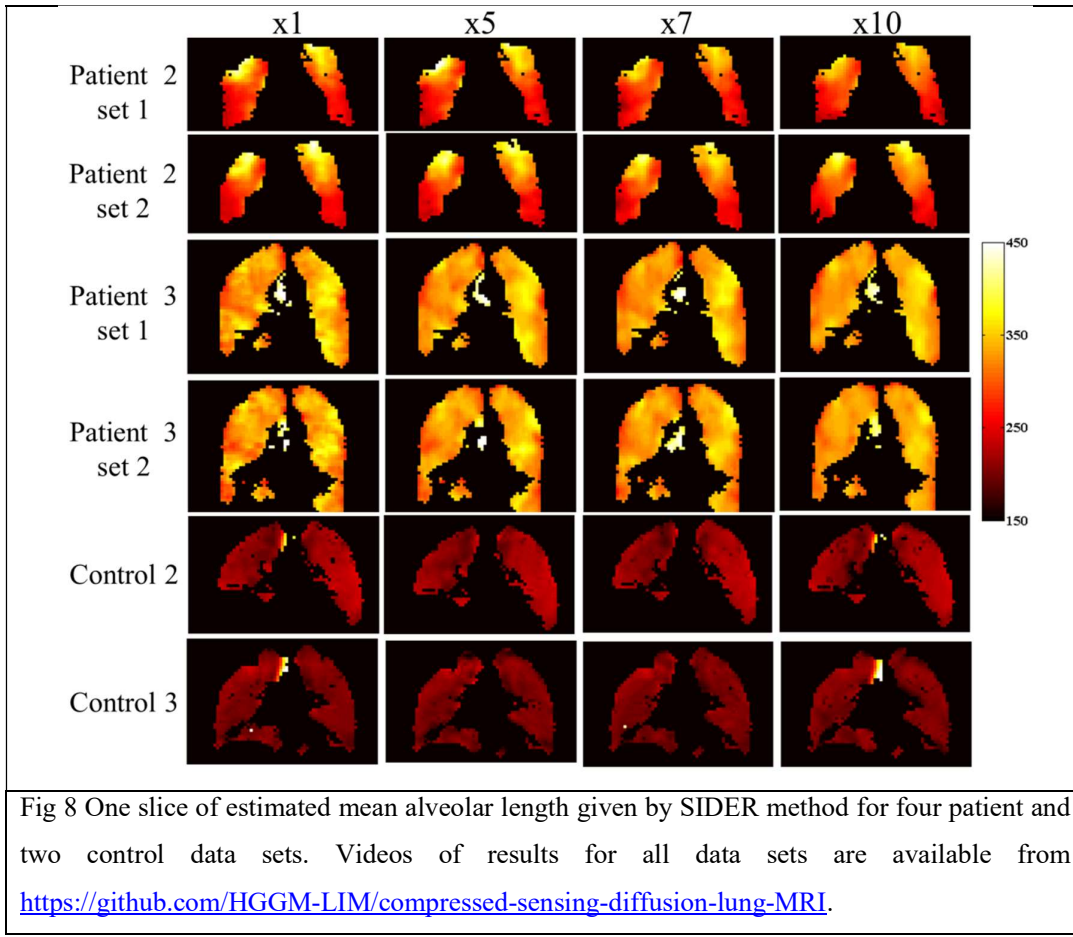

Fig 8 One slice of estimated mean alveolar length given by SIDER method for four patient and two control data sets. Videos of results for all data sets are available from https://github.com/HGGM-LIM/compressed-sensing-diffusion-lung-MRI.

Fig 9 shows histograms of estimated mean alveolar length obtained from fully sampled data using the SIDER method for acceleration factor x7 for all datasets. Histograms were very close to those of fully sampled data and variations derived from the undersampling were much smaller than intragroup differences in control and patient datasets.

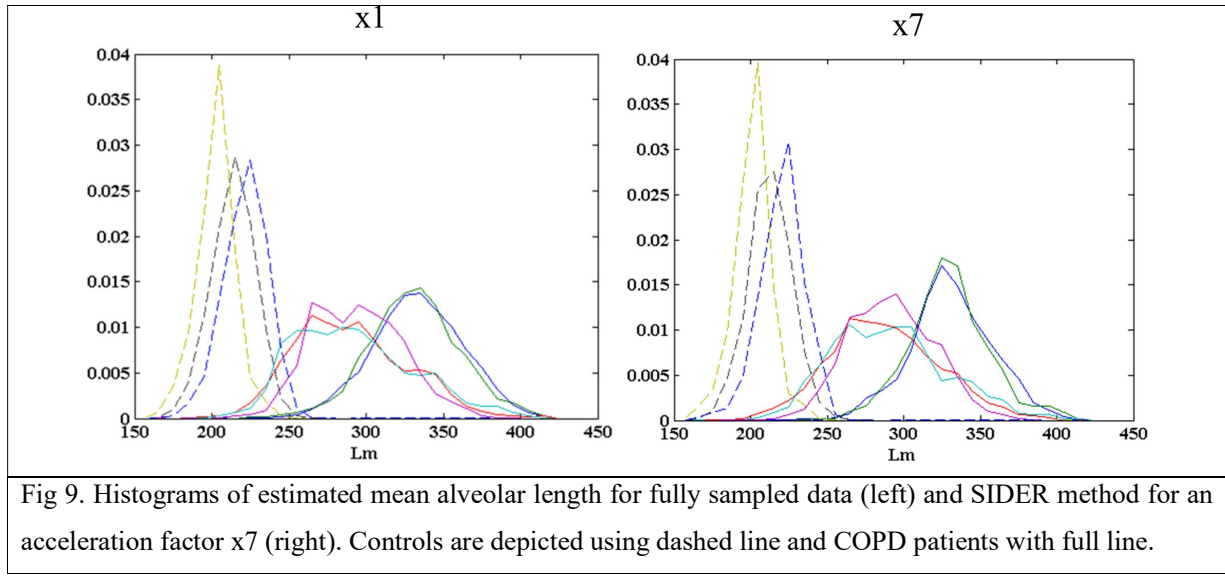

Fig 9. Histograms of estimated mean alveolar length for fully sampled data (left) and SIDER method for an acceleration factor x7 (right). Controls are depicted using dashed line and COPD patients with full line.

## Computation time

The code was implemented in MATLAB on a Windows computer with 64-bit operating system, i7-3770, 3.40 GHz CPU and 16 GB RAM. SIDER took 28 s to reconstruct all ventilation images (for 5 values of b) for one slice. We used a straightforward parallelization of SIDER over the five slices to reduce computation time.

## Discussion

We have proposed and validated a novel compressed sensing method that incorporates a model of the signal decay into the reconstruction method as prior information. The proposed method, SIDER, has been validated on both control and COPD patient data (n=8). Using retrospective undersampling, we found that accelerations of at least x7 are achievable with negligible effect on the estimates of ventilation images and estimated mean alveolar length maps. This acceleration factor (x7) is very relevant as it could be used to increase the resolution by two-fold and at same time in the x-y plane, in the number of slices and in the number of b-values. This would not only increase the resolution and volume coverage, but also improve the accuracy of estimated microstructural dimensions and may enable the use of models with a larger number of parameters.

The claimed acceleration factor depends on the criterion chosen to decide if a result is acceptable. Errors remained small and histograms were similar to those obtained from fully sampled data for acceleration factors up to x10. Image quality was preserved with small deviations in the estimated maps at a factor x10. For higher accelerations, errors were larger and the number of encoding lines acquired was very small for the present resolution (for images of

64x64), which was not considered acceptable. An acceleration factor of x7 which would allow doubling the spatial resolution, number of slices and the number of values of b, presented negligible errors.

Previous implementations of compressed sensing achieved lower acceleration factors: x2 using spatial TV in 2D [Ajraoui 2010], x3 using spatial TV in 2D and prior knowledge of a proton image acquired during the same breathold [Ajraoui 2013], and x3 using TV in 3D [Chan 2015]. The acceleration factors achieved in this work, x7-x10, are superior, which can be explained by the exploitation of undersampling along the b-dimension and the use of a reconstruction model that incorporates prior knowledge of the signal decay. SIDER method could be also extended to the 3D case as in [Chan 2015], potentially achieving even higher accelerations.

This work is subject to several limitations. First, SIDER performance is subject to tuning of the regularization parameters. In this work we have assessed that results were consistent for a wide range of these parameters but optimal selection (in specific parameters weighting sparsity across spatial- and b-dimension) could lead to higher acceleration factors. Second, we have validated the method by retrospectively undersampling control and patient data, so implementation of the proposed compressed sensing sequence in a MR scanner will be validated in future studies. In addition, the potential gain in microstructure information with the proposed method is promising but requires a dedicated study.

In future works, this method will be extended to the acquisition of 3D diffusion data, where higher accelerations can be achieved through undersampling over three directions (e.g. 2 spatial directions plus b-value), and to other hyperpolarized gas MR applications where a model of the signal behaviour is known (e.g., $pO_2$ mapping [Marshall 2010]). Our method is not restricted to hyperpolarized gas imaging, but could also be used in other diffusion MRI applications and in metabolic imaging using hyperpolarized $^{13}C$.

In conclusion, we have validated a novel compressed sensing method for lung MRI imaging and achieved high acceleration factors, which can be used to increase the amount of data acquired during a breath-hold. This methodology is expected to improve the accuracy of estimated microstructure lung information and widen the possibilities of studying lung diseases with MRI.

## Acknowledgements

The MRI diffusion data were kindly made available by Prof. Jim M. Wild (University of Sheffield).

This project has received funding from the Universidad Carlos III de Madrid, the European Union's Seventh Framework Programme for research, technological development and demonstration under grant agreement nº 600371, Ministerio de Economía y Competitividad (COFUND2013-40258), Ministerio de Educación, cultura y Deporte (CEI-15-17) and Banco Santander; Comunidad de Madrid (BRADE-CM S2013/ICE-2958) and Instituto de Salud Carlos III (PI16/02037).

The research leading to these results has received funding from the Innovative Medicines Initiative (www.imi.europa.eu) Joint Undertaking under grant agreement n°115337, resources of which are composed of financial contribution from the European Union's Seventh Framework Programme (FP7/2007-2013) and EFPIA companies' in kind contribution.

This project has received funding from the European Union's Horizon 2020 research and innovation programme under the Marie Sklodowska-Curie grant agreement N° 701915.